\documentclass[12pt,letterpaper]{article}
\usepackage[top=1in,bottom=1in,left=1in,right=1in]{geometry}
\usepackage[utf8]{inputenc}
\usepackage[breaklinks,colorlinks,urlcolor=blue,citecolor=blue,linkcolor=blue]{hyperref}
\usepackage{xcolor}
\usepackage{graphicx}
\usepackage{wrapfig}
\usepackage{amsmath}
\usepackage{amssymb}
\usepackage{natbib}

\title{WFIRST: Enhancing Transient Science and Multi-Messenger Astronomy}
\begin{document}

%\maketitle

\Huge
\begin{center}
    WFIRST: Enhancing Transient Science and Multi-Messenger Astronomy
\end{center}
\normalsize
\bigskip
\bigskip
\bigskip

\noindent
{\bf Thematic areas:\\
3. Stars and Stellar Evolution\\
4. Formation and evolution of compact objects\\
7. Cosmology and Fundamental Physics\\
8. Multi-Messenger Astronomy and Astrophysics}\\

\noindent
{\bf Principal Author:}\\
Ryan~J.~Foley\\
UC Santa Cruz\\
foley@ucsc.edu

\bigskip

\noindent
{\bf Co-authors:}\\
Joshua S.\ Bloom (UC Berkeley), S.\ Bradley Cenko (NASA GSFC), Ryan Chornock (Ohio U), Georgios Dimitriadis (UC Santa Cruz), Olivier Dor\'e (JPL), Alexei V.\ Filippenko (UC Berkeley), Ori D.\ Fox (STScI), Christopher M.\ Hirata (OSU), Saurabh W.\ Jha (Rutgers), David O.\ Jones (UC Santa Cruz), Mansi Kasliwal (Caltech), Patrick L.\ Kelly (Minnesota), Charles D.\ Kilpatrick (UC Santa Cruz), Robert P.\ Kirshner (Moore/Harvard), Anton M.\ Koekemoer (STScI), Jeffrey W.\ Kruk (NASA GSFC), Kaisey S. Mandel (Cambridge), Raffaella Margutti (Northwestern), Vivian Miranda (Arizona), Samaya Nissanke (GRAPPA/Amsterdam), Armin Rest (STScI), Jason Rhodes (JPL), Steven A.\ Rodney (South Carolina), Benjamin M.\ Rose (STScI), David J.\ Sand (Arizona), Daniel M.\ Scolnic (Duke), K.\ Siellez (UC Santa Cruz), Nathan Smith (Arizona), David N.\ Spergel (Princeton/CCA), Louis-Gregory Strolger (STScI), Nicholas B.\ Suntzeff (TAMU), Lifan Wang (TAMU), Edward J.\ Wollack (NASA GSFC)

\bigskip

\noindent
{\bf Abstract:}\\
Astrophysical transients have been observed for millennia and have shaped our most basic assumptions about the Universe.  In the last century, systematic searches have grown from detecting handfuls of transients per year to over 7000 in 2018 alone.  As these searches have matured, we have discovered both large samples of ``normal'' classes and new, rare classes.  Recently, a transient was the first object observed in both gravitational waves and light.  Ground-based observatories, including LSST, will discover thousands of transients in the optical, but these facilities will not provide the high-fidelity near-infrared (NIR) photometry and high-resolution imaging of a space-based observatory.  {\it WFIRST} can fill this gap.  With its survey designed to measure the expansion history of the Universe with Type Ia supernovae (SNe~Ia), {\it WFIRST} will also discover and monitor thousands of other transients in the NIR, revealing the physics for these high-energy events.  Small-scale GO programs, either as a supplement to the planned survey or as specific target-of-opportunity observations, would significantly expand the scope of transient science that can be studied with {\it WFIRST}.

\pagenumbering{gobble}
\newpage
\pagenumbering{arabic}

\section{Transient Science}

Exploding stars and other astrophysical transients have led to major discoveries in a wide range of astrophysical studies.  Besides being intrinsically interesting, they affect almost every aspect of astronomy.  In particular, astrophysical transients:
\begin{itemize}
  \setlength\itemsep{0pt}
    \item Are the endpoints of stellar evolution.
    \item Produce essentially all of the heavy elements in the Universe.
    \item Can produce detectable gravitational waves and neutrino emission.
    \item Reveals the rate of star formation.
    \item Heat and shape the interstellar medium.
    \item Generate significant amounts of dust in their remnants.
    \item Affect galaxy formation and evolution.
    \item Create compact objects including neutron stars and black holes.
    \item Accelerate cosmic rays in their remnants.
    \item Backlight interstellar and intergalactic media.
    \item May be the best way to detect Population III stars.
    \item Used as standard candles, can be used to measure the expansion rate of the Universe and constrain the properties of dark energy.
\end{itemize}

The 2010 Decadal Survey Report, {\it New Worlds, New Horizons} (NWNH), identified transient science being the confluence of several ``Discovery'' areas ``On the Threshold.''  In particular, of the five identified areas, transient science directly impacts the ``time-domain astronomy'' and ``gravitational wave astronomy'' categories.  Since that report, the number of transients discovered (and reported) per year has increased by a factor of 14 (from 586 in 2010 to 8382 in 2018; e.g., \citealt{Gal-Yam13}).  In 2017, the first electromagnetic (EM) counterpart \citep{Abbott17:mma} to a gravitational wave (GW) source \citep{Abbott17:gw170817}, a transient object called a ``kilonova,'' was discovered \citep{Coulter17}.  Clearly, {\bf transient astronomy has crossed the threshold and now represents one of the fastest-growing and scientifically interesting astronomical subfields.}

Transient astronomy also touched on several other NWNH themes, including the following questions:
\begin{itemize}
  \setlength\itemsep{0pt}
    \item What were the first objects to light up the Universe and when did they do it?
    \item How do stars and black holes form?
    \item How do massive stars end their lives?
    \item What are the progenitors of SNe~Ia and how do they explode?
    \item Why is the Universe accelerating?
    \item What controls the masses, spins, and radii of compact stellar remnants?
\end{itemize}
While much progress has been made on these questions, none has been sufficiently answered.

Larger samples will provide more statistics to find subtle trends in common classes of transients, identify extreme examples of these well-known classes, and discover new, rare classes of transients.  In the last decade, the increased discovery rate has led to, for instance, studies of the luminosity dependence of SNe~Ia with their local star-formation rate \citep[e.g.,][]{Jones18}, the discovery of a SN~II that persisted for more than a year \citep{Arcavi17}, and the definition of a new class of thermonuclear SNe \citep[SNe~Iax;][]{Foley13:iax, Jha17}.

Discovering transients at redshifts higher than in current samples will constrain progenitor scenarios and further connect stellar death to star formation.  SN rates determined from high-redshift samples obtained with {\it HST} represent some of the best constraints on core-collapse SN and SN~Ia progenitors \citep{Graur14, Rodney14, Strolger15}; larger samples extending to higher redshift will further improve these constraints.

Observations in the near-infrared (NIR) would enable additional opportunities, allowing for study of transients embedded in dusty environments and additional data beyond typical observations obtained today.  Finally, transients observed from space will naturally generate large samples of local environments of these objects, further connecting the explosions to the stars that produced them.

New technologies and facilities in the 2020s, and especially {\it WFIRST}, will continue the exponential progress we have achieved over the last decade.

\section{WFIRST}

{\it WFIRST} is the top-ranked space experiment from NWNH.  The current design is a 2.4-m telescope with a coronagraph and a $0.282\deg^2$ field-of-view instrument.  The wide-field instrument, which will be more important for transient science, contains seven wide filters spanning roughly 0.5--2.0~$\mu$m, an $R \approx 600$ grism (spanning 1.0--1.9~$\mu$m), and an $R \approx 100$ prism (spanning 0.6--1.8~$\mu$m).

The mission is expected to last at least 5 yr with several surveys planned.  For transient science, the most important are the SN survey (lasting 2 yr and a total of 6 months of telescope time) and the general observer (GO) program (about 25\% of telescope time).

A more complete summary can be found in a recent white paper \citep{Akeson19}.  Additional, albeit sometimes outdated, information can be found in \citet{Spergel15}.

\section{The WFIRST Supernova Survey}

\citet{Hounsell18} presented realistic simulations of several potential {\it WFIRST} SN surveys.  Although these are not optimized and several are made obsolete by the deletion of the Integral Field Channel, the simulations are sufficient for planning purposes.

We will focus on the ``Imaging:Allz'' survey.  It consists of three imaging tiers achieving an individual exposure depth (total stacked depth) of 22.3, 24.5, and 26.1~mag (25.0, 27.2, and 28.8~mag) in $RZY\!\!J$, $RZY\!\!J$, and $Y\!JH\!F$ covering 49, 20, and 9~deg$^{2}$ for each tier, respectively.  The survey has a five-day cadence over the middle two years of the five-year mission.  This survey would measure useful distances to $\sim$11,000 SNe~Ia to $z \approx 3$.  The Imaging:Allz survey uses the entire 6 months of survey time for imaging; however, with the recent addition of the prism to WFIRST's hardware, it is likely that prism observations will account for 10--30\% of the survey time, reducing the total number of SNe and the survey area by a similar fraction.

\begin{wrapfigure}{r}{0.5\textwidth}
 \vspace{-25pt}
  \begin{center}
    \includegraphics[width=0.48\textwidth]{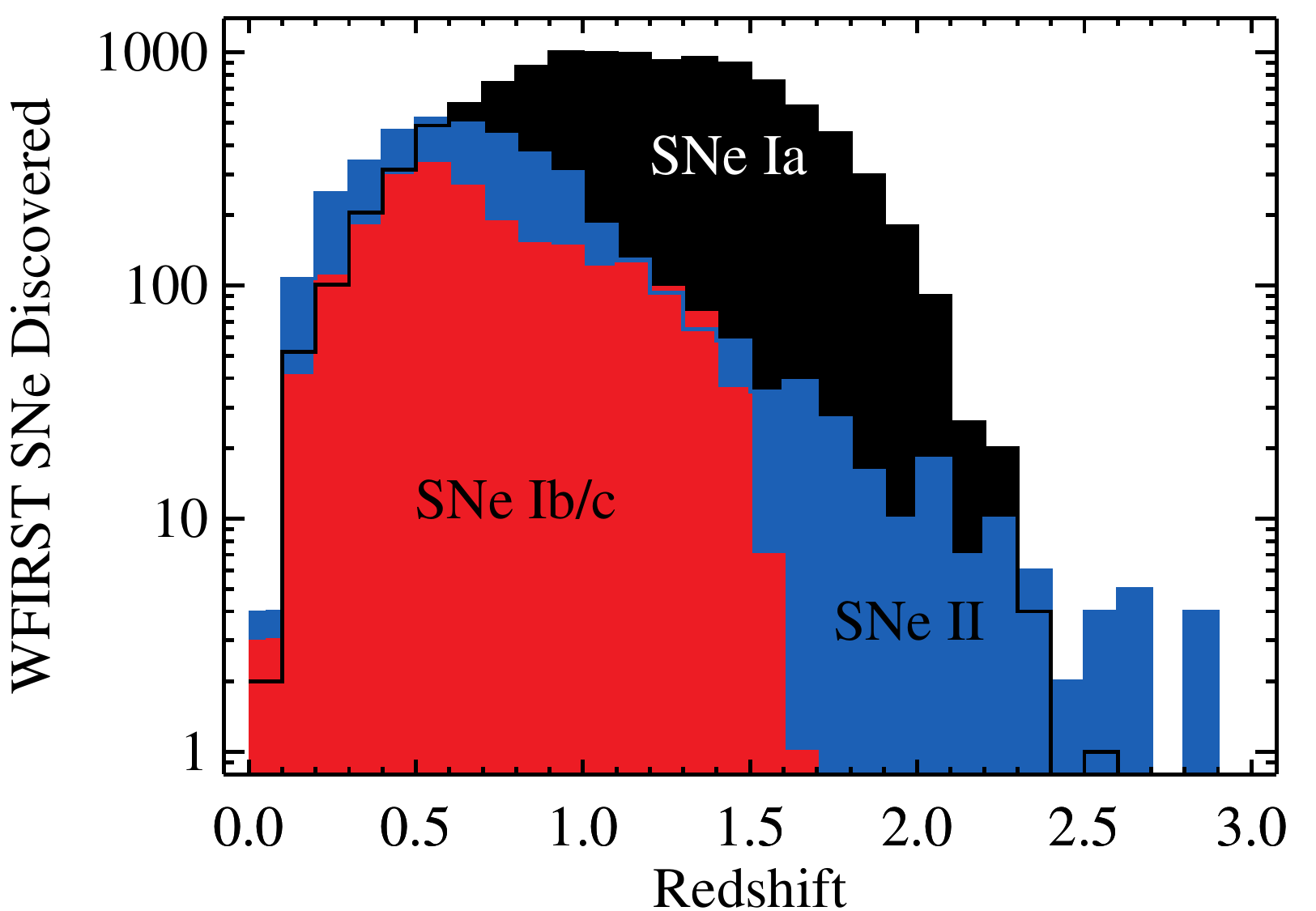}
  \end{center}
 \vspace{-20pt}
 \caption{Redshift distributions for a possible WFIRST SN survey.  The black, blue, and red histograms represent SNe~Ia, SNe~II (SNe~IIP, SNe~IIL, and SNe~IIn), and SNe~Ib/c (SNe~Ib and SNe~Ic), where at least one data point has ${\rm S/N} \ge 5$.  There are a total of 11,504, 2283, 1077, 644, 1466, and 702 SNe~Ia, IIP, IIL, IIn, Ib, and Ic, respectively.}\label{f:hist}
\vspace{-15pt}
\end{wrapfigure}

To fully exploit the data from the {\it WFIRST} SN survey, we must look beyond only the cosmologically useful SNe~Ia.  \citet{Hounsell18} performed detailed simulations of possible {\it WFIRST} SN surveys, providing a baseline for additional investigations.  As an example, the three-tier, four-band survey is expected to discover $\sim$11,500 SNe~Ia with a photometric signal-to-noise ratio (S/N) of $\ge$5 in at least one band at one epoch.  This survey would discover SNe~Ia to at least a redshift of 3.

This same survey would discover $\sim$2300 (9100) SNe~II and $\sim$650 SNe~Ib/c.  A histogram of the expected discoveries is displayed in Figure~\ref{f:hist}.  While rarer classes have not been simulated, one can estimate based on other surveys that the {\it WFIRST} SN survey should discover tens to hundreds of SNe~IIn, superluminous SNe (SLSNe), SNe~Iax, and tidal disruption events (TDEs).

\section{GO Supplements to the Supernova Survey}

The {\it WFIRST} SN survey will provide excellent light curves for transients with durations similar to that of SNe~Ia.  However, transients with shorter or longer durations may have inadequate sampling.  GO programs could leverage the investment in the SN survey to produce significant science gains with a marginal investment in telescope time.

\begin{wrapfigure}{r}{0.5\textwidth}
% \vspace{-26pt}
 \vspace{-13pt}
  \begin{center}
    \includegraphics[width=0.48\textwidth]{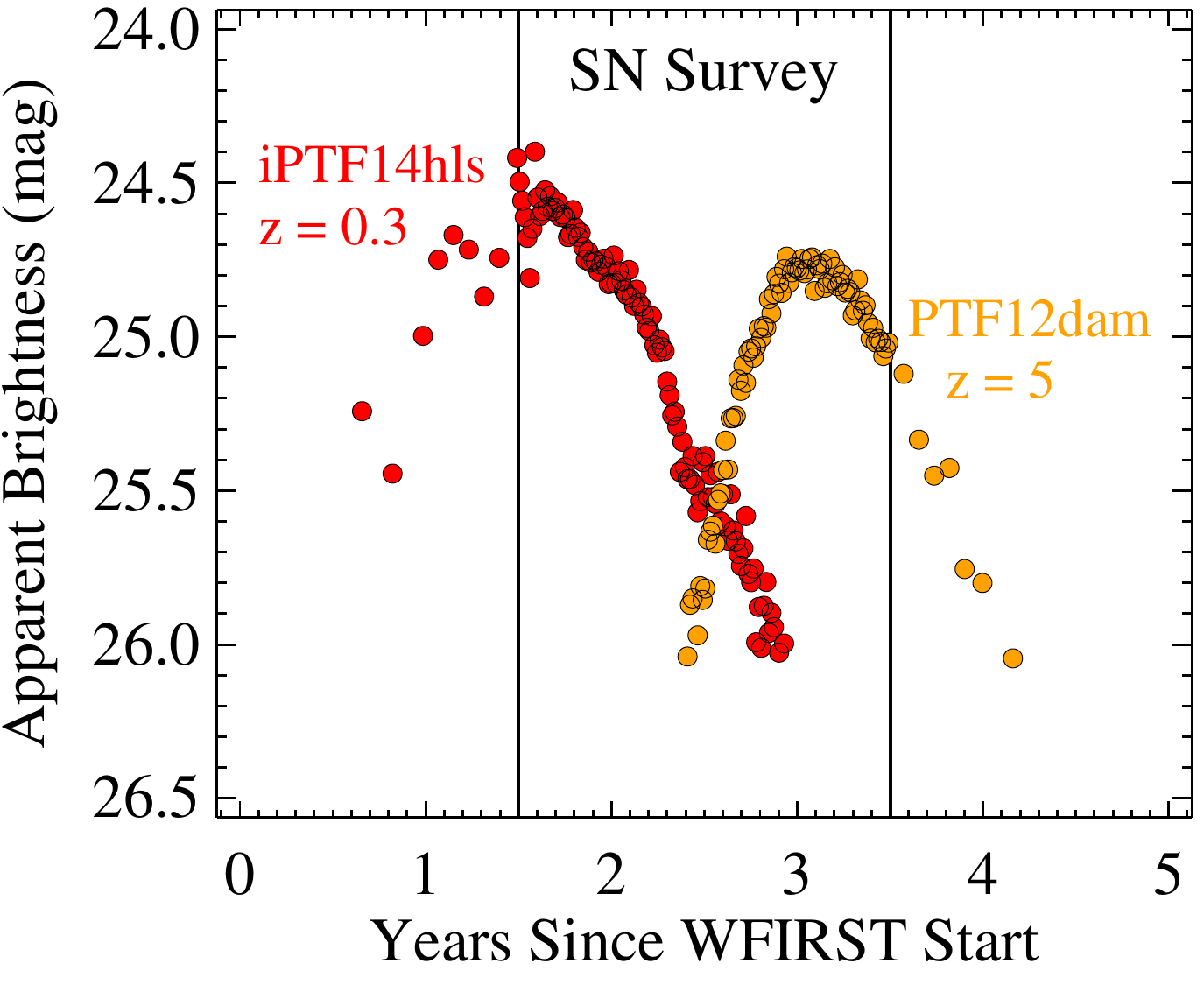}
  \end{center}
 \vspace{-20pt}
 \caption{Simulated {\it WFIRST} $H$-band light curves of SNe similar to the long-lived SN~II iPTF14hls \citep{Arcavi17} at $z = 0.3$ in red and the SLSN PTF12dam at $z = 5$ \citep{Nicholl13} if they were observed in the Deep SN tier with arbitrary explosion times. The middle two years corresponding to the {\it WFIRST} SN survey are marked.  Observations from a possible GO program with a cadence of 2 months in the first and last year of the mission and every 2 weeks for the 6 months before/after the SN survey are included to indicate how additional observations enhance the science.}\label{f:lc}
%\vspace{-15pt}
\vspace{-10pt}
\end{wrapfigure}

Some of the most interesting transients have intrinsically long durations.  For instance, some SLSNe and TDEs can be persist for months or even years \citep[e.g.,][]{Gezari15, Dong16, Arcavi17}.  Moreover, very high-redshift SNe will have their light curves time dilated by the factor $1+z$.  {\it WFIRST} has the ability to detect some SNe at $z = 10$--20 \citep{Tanaka12, Whalen13:pi, Whalen13:iin}, where a month in the SN's reference frame would correspond to more than a year in the observer frame.  While a subset of these long-lasting events may have sufficient light curves from the SN survey alone, most would suffer from the edge effects of the survey resulting in incomplete light curves (Figure~\ref{f:lc}).

A simple GO survey that monitors the SN fields, perhaps monthly or weekly right before/after the SN survey and at longer timescales further from the SN survey, would provide the necessary coverage to complete most light curves of long-duration transients observed during the SN survey.

Since the SN fields will be deep fields for extragalactic astronomy and calibration fields for the observatory, other science programs would also benefit from such a program.  Moreover, long-term monitoring of the SN fields for variable targets as well as preparation for the SN survey would be extremely beneficial.

Some transients, both theoretical and observed, have timescales significantly shorter than a SN~Ia \citep[e.g.,][]{Foley09:08ha, Drout14, Siebert17, Margutti19, Perley19}, and a 5-day cadence survey may result in only a few data points for those objects, if they are detected at all.  A GO program that fills in the gaps between normal SN survey observations would be more economical than a stand-alone survey.

Finally, GO programs to either supplement the SN survey with additional filters or perhaps deeper exposures in a given epoch could benefit particular science cases.  For instance, there will likely be an ``ultra-deep'' survey focusing on a single {\it WFIRST} pointing at the center of a SN deep field.  If observations are spread out temporally, this 0.25-deg$^{2}$ field --- 81 times larger than the {\it Hubble} Ultra-Deep Field --- will produce a smaller number of SNe, but will probe lower-luminosity SNe at low redshift and likely find the highest-redshift SNe.

\section{WFIRST and Multi-Messenger Astronomy}

With the discovery of the first EM counterpart to a GW source, we now have confirmed theoretical predictions that neutron-star mergers should produce a radioactively powered kilonova \citep[e.g.,][]{Li98, Rosswog02, Metzger10, Kasen13}.  While the kilonova associated with GW170817 was initially blue, the blue component (likely a physically distinct component) faded on the timescale of days.  On the other hand, the kilonova peaked in the NIR a few days after discovery and faded slower \citep[e.g.,][]{Drout17}.  While the luminosity of the blue component should depend strongly on the physical conditions of the merger and our viewing angle, the red component is expected to be more ubiquitous, consistent, and isotropic.

In many scenarios, and especially for faint/red kilonovae, {\it WFIRST} may even be the only observatory capable of discovering the EM counterpart, and thus the precise position of the GW source.  The exquisite wide-field mapping speed,  red sensitivity, and (potential) rapid response capability of {\it WFIRST} are well-suited to this search.  As localization regions continue to decrease, especially with the incorporation of KAGRA and LIGO India into the network, a larger fraction of GW events will be localized to $<$10~deg$^{2}$, where searching with {\it WFIRST} is practical \citep{Nissanke13}.  Notably, the first kilonova was discovered by the Swope telescope and its 0.25-deg$^{2}$ camera \citep{Coulter17}.

Such a search would require relatively fast turn-around target-of-opportunity capabilities with an execution time of at most a few days\footnote{The current design requirements include target-of-opportunity observations, but with a delay of up to 2 weeks after approval --- too long for this science case.  The current system is capable of faster responses, and the delay will likely be limited by policy.}.  An example strategy is presented in Appendix A-52 of \citet{Spergel15}.

Even in cases where another facility discovers an EM counterpart, {\it WFIRST} will be critical for long-term monitoring, including spectroscopy with the prism and grism.

Additionally, simulations by \citet{Scolnic18:kn} found that the {\it WFIRST} SN survey has the potential to independently discover $\sim$8 kilonovae per year to $z = 0.8$, a discovery rate higher than that of LSST. However, the 5-day cadence is  generally insufficient for unambiguous identification and detailed astrophysical characterization.  Supplementing the SN survey to obtain higher-cadence observations would likely increase the overall yield as well as improve constraints on derived physical parameters.  Future gravitational wave experiments can take advantage of these higher-redshift discoveries with advanced LIGO plus, and the proposed Cosmic Explorer and Einstein Telescope (with horizon distances out to $z = 0.5$, which are beyond the reach of even LSST for discovery).

\section{Summary}

While much progress has been made over the last decade in discovering, characterizing, and understanding astrophysical transients, the fundamental questions posed by NWNH related to transients remain mostly unanswered.  To make progress on these important questions and newly formed questions, we must obtain observations that push into new dimensions.  While ground-based surveys, such as LSST, will discover thousands of transients, {\it WFIRST} has the unique ability to provide a large sample with high-quality optical through NIR data to faint flux levels and with high-resolution imaging of their environments.  The resulting dataset will be generated primarily as a byproduct of the already planned SN survey to study dark energy.  Minimal additional observations including those taken outside the nominal SN survey and additional epochs during the SN survey, as well as target-of-opportunity capabilities, would greatly enhance the value of the mission.

\newpage

\bibliographystyle{yahapj}
\bibliography{decadal}

\end{document}